# Survey on security issues in file management in cloud computing environment

Udit Gupta
Department of Computer Science and Information Systems
Birla Institute of Technology and Science, Pilani – Goa Campus
India
uditg.89@gmail.com

**ABSTRACT**
Cloud computing has pervaded through every aspect of Information technology in past decade. It has become easier to process plethora of data, generated by various devices in real time, with the advent of cloud networks. The privacy of user's data is maintained by data centers around the world and hence it has become feasible to operate on that data from lightweight portable devices. But with ease of processing comes the security aspect of the data. One such security aspect is secure file transfer either internally within cloud or externally from one cloud network to another. File management is central to cloud computing and it is paramount to address the security concerns which arise out of it. This survey paper aims to elucidate the various protocols which can be used for secure file transfer and analyze the ramifications of using each protocol.

## 1. INTRODUCTION

Ever since internet became a force in the field of communication, there has been an increasing demand in the storage and security of data. This demand grew exponentially in the last two decades which resulted in the advent of cloud computing [1, 12]. The issue pertaining to storage of data has been addressed to a great extent but security still remains a challenge for the researchers as has been highlighted in [2, 3, 13 and 14]. Furthermore, in case of wireless networks the problem pertaining to security has only aggravated with the possibility of sleep deprivation attacks [17] becoming inevitable. File transfer between virtual machines (wired or wireless) or from one cloud network to another is one such aspect which needs to be addressed as part of cloud security. File transfer protocol (FTP) [5] is generally used for transferring files but there are number of security vulnerabilities with FTP in its raw form as has been described in RFC 2577 in [4]. The purposes of FTP does not include any mention of security and hence FTP has been exploited innumerable times. All transmissions are in plain text including usernames and passwords and hence anyone with proper configuration can perform packet sniffing to obtain sensitive information on the network.

## 2. SECURE PROTOCOLS FOR FILE TRANSFER

To address the vulnerabilities in FTP, developers have come up with various solutions which involve integrating FTP with secure protocols like SSH [10], SSL [6, 7, 8 and 9] or simply use scp (secure copy) to copy files between machines. This paper will describe several such protocols that will take into account the confidentiality and integrity of data. Furthermore, performance of these protocols will also be discussed which will include how encryption overheads might impact the overall operation in terms of delay or resource usage.

## 2.1 Secure shell file transfer protocol (SFTP)

SFTP is responsible for accessing and managing files on remote systems using SSH. The major difference between SFTP and FTP is that the former encrypts commands and data both, preventing sensitive information from being transmitted in clear text over the network. SFTP can perform number of operations on remote machines like obtaining file size, traversing through directories, counting number of files in a directory, removing files, creating symbolic link and many more. Due to the ability of SFTP to piggy back on SSH connection, the overhead incurred during encryption will not differ much from SSH.

SFTP is also able to transfer files between remote machines having different operating systems. For instance, in order to transfer files to windows OS, manual installation of SSH server needs to be performed on windows. Since IIS (internet information services) from Microsoft only provides support for ftp, there are SSH servers like winsshd, freesshd, freeftpd, filezilla [11] which can be used for SFTP support.

The authentication via SFTP can be performed in two ways: password authentication and public key authentication. Both these methodologies are equally secure but latter would take less time. On the other hand, key management might become an issue if there are too many keys in which case password based authentication would take precedence.

## 2.2 File transfer protocol over Secure Socket layer (FTPS)

FTPS is also responsible for accessing and managing files on remote systems. The major difference as compared to SFTP however lies in the underlying protocol being used for adding security. In case of FTPS the underlying protocol used is SSL/TLS. The FTPS protocol uses X.509 certificates for authentication which in turn contains the public key and information about the certificate owner.

There are two different forms of FTP over SSL: Explicit and Implicit encryption. In case of explicit encryption, connection begins with a clear text and becomes encrypted after an AUTH command is sent while in case if implicit encryption is starts off encrypted. For FTPS support provided by most platforms the default setting is explicit encryption.

## 2.3 SSH Tunneling

A Secure Shell (SSH) tunnel is an encrypted tunnel based on SSH protocol connection through data can be transferred. The data via this tunnel is generally transmitted in an unencrypted format. OpenSSH software is used to establish tunnel between two remote machines. SSH Tunneling can be used to transfer files securely but it is generally recommended for overcoming firewall settings.

Port forwarding is used to direct all the traffic to a remote machine. There are two kinds of port forwarding which are used: local port forwarding and remote port forwarding. In local port forwarding we define a local port which we then use it to transfer files to remote machine by forwarding this port. In remote port forwarding, the application on remote machine can access services on local host.

## 2.4 SCP and RSYNC protocol

Secure copy (SCP) is another network protocol which is used for file transfer between remote hosts. The underlying protocol responsible for providing authentication is SSH. RSYNC is another protocol which is used for transferring files between systems. But the major difference between SCP and RSYNC lies in the fact that former overwrites every file on the destination host while latter is more of synchronization algorithm which will perform delta transfer between the hosts. Since RSYNC only performs the delta transfer, the network resources consumed by this protocol will be much less as compared to SCP.

## 2.5 FTP over IPSec

In IPSec [15] protocol security is implemented at the network layer where packet processing takes place. Earlier protocols were implemented at the Application layer of the TCP/IP model where changes were required in individual's client machine. But in IPSec in order to enhance security, changes are required at the network layer (specifically routers) to implement it as a standard security measure across network.

IPSec itself consists of 2 main protocols: Authentication Header (AH) and Encapsulating Security Payload (ESP). ESP is the preferred choice since it provides both authentication and confidentiality while AH does not provide confidentiality protection. Due to the complexity involved and the cost incurred while implementing, IPSec is generally not used for secure file transfer. However, it is quite useful for large corporations for designing their own Virtual Private Networks (VPNs).

## 3. COMPARISON BETWEEN DIFFERENT PROTOCOLS

Table 1 provides number of seconds taken by files when getting transferred from one linux VM to another. This scenario is repeated for 3 different cases when numbers of files are 8, 100 and 500 and execution time for transfer of files is recorded in each case. The time has been recorded in seconds using the inbuilt utility called 'date' in Linux operating system and is accurate up to 2 decimal places. Also the implementation was done in an environment where both remote machines had CentOS installed and were on the same subnet so as to minimize the delay which might be incurred due to switches and routers.

As it can be seen from the table that the time taken for file transfer is least when underlying protocol is FTP since the connection is not encrypted and files are transferred in plain text due to which no encryption headers are involved. On the other hand the time taken for SFTP file transfer for $1^{st}$ case when number of files are 8 is almost 51% higher but decreases to around 35% for $2^{nd}$ and 31% for $3^{rd}$ case when number of files are 100 and 500 respectively. Given the scalability involved in cloud, we can safely assume that for thousands of files the performance degradation will come down to around 20% assuming linear regression.

When underlying protocol is FTPS, the time taken for file transfer between 2 hosts is almost similar compared to SFTP. Thus we can conclude that as far as the delay in file transfer which might be incurred on implementing FTPS or SFTP over FTP is concerned, it will almost be similar. The slight difference of the order of $10^{-2}$ seconds observed might be due to difference in network traffic on the same subnet which can vary in different situations.

In case of SCP, the time taken for file transfer between 2 hosts is comparable to what was observed for SFTP and FTPS. For 8 files the SCP time was more but for large number of files the execution time was less than what was observed for SFTP or FTPS. This might due to the fact that password authentication was used in case of SCP due to which slightly different times were observed as compared to SFTP. But since underlying protocol in SCP is also SSH so we can safely assume that in case of public key authentication performance of SCP will be similar to SFTP. However, in most scenarios SFTP is given preference since functionality of SCP is only limited to transfer of files while using SFTP we can also perform number of operations on files on remote machines.

In case of RSYNC, the time taken for file transfer is slightly less than that of SCP file transfer. This is due to the fact that RSYNC does not overwrite existing files on remote machine and performs delta operations wherein only those files are transferred which are not present in the destination directory of remote machines.

Based on the above observations it cannot however be stated assertively about which protocol should be the most ideal one. It completely depends on the particular scenario in cloud. It also depends on the operating system installed on client machines. For instance, windows do not have SSH support although higher versions of Microsoft IIS do have SSL support which in turn has better compatibility with windows. So in those cases it would be advisable to use FTPS over any other SSH based protocol.

Another factor which might be taken into consideration while choosing secure file transfer protocol is the CPU usage. Although high end servers are available in the market but due to the cost involved many small scale organizations may not be able to buy them. In that case it becomes paramount to optimize CPU usage which might go very high if thousands of files need to be transferred. In that aspect, tunneling and port forwarding techniques might come in handy. In many organizations cloud itself will be secured from external access and hence it may not be required to secure every file transfer. In this case it is better to use a protocol which has less delay time and consumes less network resources.

## 4. CONCLUSIONS

While cloud computing continues to dominate the market of the domain of Information Technology, there are number of security vulnerabilities which needs to be addressed. Some of them were highlighted in this paper and a brief attempt was made to enable the reader to understand the context in which those solutions can be deployed. Furthermore, a comparison was presented between different protocols to provide a perspective about delay incurred and network resources usage.

Table 1

|  | Time required for file transfer for different number of files | | |
|---|---|---|---|
|  | For 8 files | For 100 files | For 500 files |
| FTP | 0.35s | 1.77s | 8.16s |
| SFTP | 0.53s | 2.39s | 11.3s |
| FTPS | 0.52s | 2.35s | 11.18s |
| SCP | 0.56s | 2.0s | 10.26s |
| RSYNC | 0.55s | 1.98s | 10.15s |